\documentclass[a4paper,fleqn,usenatbib]{mnras}

\usepackage[T1]{fontenc}
\usepackage[english]{babel}
\usepackage[utf8]{inputenc}
\usepackage{ae,aecompl}

\usepackage{graphicx}	
\usepackage{float}
\usepackage{amsmath}	
\usepackage{amssymb}	
\usepackage{hyperref} %
\usepackage[dvipsnames]{xcolor}
\usepackage{lineno}

\title[Unbalanced cooling in IC\,4296]{Powerful AGN jets and unbalanced cooling in the hot atmosphere of IC 4296}

\author[R.Grossov\'a et al.]{
R. Grossov\'a,$^{1,2}$\thanks{E-mail: romana.grossova@gmail.com}
N. Werner,$^{3,1,4}$
K. Rajpurohit,$^{5}$
F. Mernier,$^{3,6,7}$
K. Lakhchaura,$^{3,8}$\newauthor
K. Gab\'anyi,$^{8,9}$
R. E. A. Canning,$^{10}$
P. Nulsen,$^{11,12}$
F. Massaro,$^{2}$
M. Sun,$^{13}$
T. Connor,$^{14}$\newauthor
A. King,$^{10}$
S. W. Allen,$^{10}$
R. L. S. Frisbie,$^{15}$
M. Donahue,$^{15}$
A. C. Fabian$^{16}$
\\
$^1$Department of Theoretical Physics and Astrophysics, Faculty of Science, Masaryk University, Kotl\'a\v{r}sk\'a 2, Brno, 611 37, Czech Republic \\
$^2$Dipartimento di Fisica, Universita degli Studi di Torino, via Pietro Giuria 1, I-10125 Torino, Italy\\
$^3$MTA-E\"otv\"os University Lend\"ulet Hot Universe Research Group, P\'azm\'any P\'eter s\'et\'any 1/A, Budapest, 1117, Hungary \\
$^4$School of Science, Hiroshima University, 1-3-1 Kagamiyama, Higashi-Hiroshima 739-8526, Japan \\
$^{5}$Dipartimento di Fisica e Astronomia, Universit\'a di Bologna, Via Gobetti 93/2, 40131, Bologna, Italy\\
$^{6}$SRON Netherlands Institute for Space Research, Sorbonnelaan 2, 3584 CA Utrecht, The Netherlands\\
$^{7}$Institute of Physics, E\"otv\"os University, P\'azm\'any P\'eter s\'et\'any 1/A, Budapest, 1117, Hungary\\
$^8$MTA-E\"otv\"os University Extragalactic Astrophysics Research Group, P\'azm\'any P\'eter s\'et\'any 1/A, Budapest, 1117, Hungary \\
$^9$Konkoly Observatory, MTA Research Center for Astronomy and Earth Sciences, Konkoly Thege Mikl\'os \'ut 15-17, H-1121 Budapest, Hungary\\
$^{10}$Kavli Institute for Particle Astrophysics and Cosmology, Stanford University, 452 Lomita Mall, Stanford, CA 94305-4085, USA \\
$^{11}$Harvard Smithsonian Centre for Astrophysics, 60 Garden Street, Cambridge, MA 02138, USA\\
$^{12}$ICRAR, University of Western Australia, 35 Stirling Hwy, Crawley,
WA 6009, Australia\\
$^{13}$Department of Physics and Astronomy, University of Alabama in Huntsville, Huntsville, AL 35899, USA\\
$^{14}$The Observatories of the Carnegie Institution for Science, 813 Santa Barbara Street, Pasadena, CA 91101, USA\\
$^{15}$Physics $\&$ Astronomy Department, Michigan State University, East Lansing, MI 48824-2320, USA\\
$^{16}$Institute of Astronomy, University of Cambridge, Madingley Road, Cambridge CB3 0HA, UK\\
}
\date{Accepted 2019 June 19. Received 2019 June 19; in original form 2019 March 5}

\pubyear{2019}

\begin{document}
\label{firstpage}
\pagerange{\pageref{firstpage}--\pageref{lastpage}}
\maketitle


\begin{abstract}
We present new Karl G. Jansky Very Large Array (VLA, 1.5\,GHz) radio data for the giant elliptical galaxy IC\,4296, supported by archival radio, X-ray (\textit{Chandra}, \textit{XMM-Newton}) and optical (SOAR, {\it HST}) observations. The galaxy hosts powerful radio jets piercing through the inner hot X-ray emitting atmosphere, depositing most of the energy into the ambient intra-cluster medium (ICM). Whereas the radio surface brightness of the A configuration image is consistent with a Fanaroff-Riley Class I (FR\,I) system, the D configuration image shows two bright, relative to the central region, large ($\sim 160\,\rm{kpc}$ diameter), well-defined lobes, previously reported by Killeen et al., at a projected distance $r\gtrsim 230\,\rm{kpc}$. The \textit{XMM-Newton} image reveals an X-ray cavity associated with one of the radio lobes.
The total enthalpy of the radio lobes is $\sim 7 \times10^{59}\,\rm{erg}$ and the mechanical power output of the jets is $\sim 10^{44}\,\rm{erg\,s}^{-1}$. 
The jets are mildly curved and possibly re-brightened by the relative motion of the galaxy and the ICM. The lobes display sharp edges, suggesting the presence of bow shocks, which would indicate that they are expanding supersonically.
The central entropy and cooling time of the X-ray gas are unusually low and the nucleus hosts a warm H$\alpha +$[\ion{N}{II}] nebula and a cold molecular CO disk. Because most of the energy of the jets is deposited far from the nucleus, the atmosphere of the galaxy continues to cool, apparently feeding the central supermassive black hole and powering the jet activity.

\end{abstract}




\section{Introduction}
\label{sect:intro}
The oldest, largest known galactic structures in the Universe are giant elliptical galaxies. These systems are evolving contrary to the standard hierarchical galaxy formation scenario \citep{thomas2002}. Observations and theoretical models indicate that they formed around 0.2--1~billion years after the Big Bang, went through a fast star-forming phase, and afterward grew only by dry mergers \citep{thomas2010,onodera2015}.

Initially, the evolution of giant ellipticals is dominated by dark matter, which clumps into haloes via gravity. When the mass of a halo reaches $\sim 10^{12}\,M_{\odot}$ \citep{correa2018}, the inflowing gas passes through accretion shocks, and its temperature increases to several million Kelvin, forming an X-ray emitting atmosphere \citep[further replenished by stellar mass loss;][]{pellegrini2018} and quenching star-formation \citep{cattaneo2009}. These galaxies will continue to grow passively by interactions and dry mergers with other galaxies \citep{cattaneo2009}. On the other hand, the sub-Eddington accretion rate of the gas onto the central supermassive black hole enables the so-called radio-mechanical (or radio mode) active galactic nucleus (AGN) feedback, thereby producing jets and lobes of relativistic plasma, visible in the radio band, which are able to propagate well outside their galaxy hosts. 

Radio wavelength studies of giant elliptical galaxies are important for answering outstanding questions about the heating and cooling of their X-ray emitting atmospheres and about the physics of their AGNs \citep{brighenti2006}. In clusters of galaxies, there is strong observational evidence for radio-mode AGN feedback balancing the cooling of the intra-cluster medium \citep[ICM; e.g][]{churazov2000,fabian2003,birzan2004,rafferty2006,bruggen2002} and preventing cooling flows and dramatic star formation \citep[see][]{fabian1994}. It is believed that radio-mechanical feedback plays a critically important role also at smaller scales, in preventing atmospheric cooling in massive galaxies \citep{werner2019}. \ 

In the case of giant ellipticals, the so-called cooling flow problem is more severe. Shorter cooling time and larger amount of gas returned by the stellar populations over the lifetime of the galaxy \citep{cattaneo2009} place stronger demands on the AGN feedback as a source of heat balancing the radiative cooling. In some cases, the high-velocity radio jets are able to drill through the hot galactic atmospheres without significantly affecting the host galaxy, which may lead to unbalanced cooling in the innermost part of the galaxy \citep{sun2005}. 
Comprehensive studies of such systems are essential for our understanding of the role of AGN feedback in galaxies, galaxy groups, and galaxy clusters.

The powerful AGN jets penetrating the innermost hot galactic atmosphere and depositing most of their energy out at radii $r\gtrsim230\,\rm{kpc}$ make the giant elliptical galaxy IC\,4296, in the central region of the galaxy cluster Abell~3565, a source of high interest. Its radio counterpart, PKS~1333-33, has attracted the attention of astronomers for over 30~years. Historically, the first comprehensive studies were published by \cite{killeen1986I,killeen1986II} and \cite{killeen1988} using data from Very Large Arrray (VLA) at 1.3, 2, 6, and 20\,cm. In particular, \cite{killeen1986II} analysed X-ray data of the system from the imaging proportional counter (IPC) onboard the {\it Einstein} observatory. In addition to prominent jets extending over 35\,arcmin, the authors reported a barely resolved diffuse X-ray source coinciding with the position of the galaxy.

In this paper, we present new radio data of IC\,4296 obtained with the VLA in the A and D configurations at 1.5\,GHz and compare them with archival observations available at lower frequencies in the TIFR GMRT Sky Survey \citep[TGSS;][]{intema2017} and in the GaLactic and Extragalactic All-sky Murchison Widefield Array \citep[GLEAM;][]{hurley-walker2016}.

While our paper presents new, deep radio observations, our goal is to study this system in a more comprehensive way. Therefore, we also analyse archival X-ray data from the {\it Chandra} and {\it XMM-Newton} observatories, as well as narrow-band H$\alpha +$[\ion{N}{II}] images from the 4.1 meter Southern Astrophysical Research (SOAR) telescope and optical data from the Hubble Space Telescope ({\it HST}). This allows us to study the interaction between the jet/lobes and the surrounding hot and cold thermal gas phases.\

Our paper is organized as follows: Section\,\ref{sect:observations} includes the observational details and data reduction methods.
Our results are presented and described in Section\,\ref{sect:results}. A thorough discussion in Section\,\ref{sect:discussion} addresses the most important outcomes of our multi-wavelength analysis connecting all these multi-wavelength observations, with a particular emphasis on the relation between the radio and X-ray band. Finally, our conclusions and remarks are summarized in Section\,\ref{sect:summary}.\

Throughout this paper, we used the following cosmological parameters: $\rm H_{0}=69.6\,\rm{km\,s}^{-1}\,\rm{Mpc}^{-1}$ \citep{bennett2014}, $\Omega_\mathrm{M}=0.286$ and $\Omega_{\Lambda}=0.714$.\
 At the redshift of 0.0125 the distance of IC\,4296 is $49\,\rm{Mpc}$ \citep{mei2000}, estimated using the~surface brightness fluctuation method, with the corresponding scale of $0.256\,\rm{kpc}/\rm{arcsec}$.\
 The spectral index $\alpha$ of the synchrotron radiation as a function of the flux density $S_\nu$ at the frequency $\nu$ is defined as $S_\nu \propto \nu^{\alpha}$.\
 The elemental abundances are expressed with respect to the proto-solar values from \citet{lodders2009}.

\section{Observations and data analysis}
\label{sect:observations}

\subsection{Radio observations and analysis}
\label{sect:radio_observations}
IC\,4296 was observed on 2015 July 11 by the Karl G. Jansky VLA in A configuration (project code: 15A-305; PI: Werner) and on 2018 October 26 (project code: 18A-317, PI: Grossova) in D configuration. The L-band receiver covering the frequency range spanning from 1 to 2\,GHz is divided into sixteen spectral windows with 64 1000\,kHz wide channels. The one-hour long observations consist of $\approx 40$\,minutes integration time on the target and $\approx 10$\,minutes on both the standard VLA flux density calibrator 3C\,286 and a~nearby compact source J1316-3338 used as amplitude and phase calibrator.
The VLA data were reduced and imaged using the National Radio Astronomy Observatory ({\sc NRAO}) pipeline Common Astronomy Software Applications \citep[{\sc casa},][]{mcmullin2007}, v4.7.2 and v5.0.0 for A and D configuration data, respectively. 

A careful approach was followed to flag Radio Frequency Interferences (RFI). First, we used the auto-flagging algorithm {\tt tfcrop} in {\sc casa} to identify the outliers in the time-frequency plane. Then, we used the Offringa's RFI software {\sc aoflagger} \citep{offringa2012}. The combination of the two strategies, for the A configuration, flagged in total 46\,$\%$ of the data, with spectral windows 2, 3, 8 and 9 almost entirely flagged. In case of the D configuration, more than half of the data had to be flagged. 

Afterwards, we used a model for the flux calibrator 3C\,286 provided by the {\sc casa} package to set up the flux density scale determined by \cite{perley2013}. The initial gain calibration was performed for central channels on both calibrators to average over the small variations of phase with time in the bandpass. Next, we performed the bandpass calibration. The relative delays of each antenna in comparison with the reference antenna were derived and complex bandpass solutions were calculated to include variations in gain with frequency. The next step determined complex gains for both calibrators. Finally, the derived calibration solutions were applied to the target.

The deconvolution and imaging were performed by the {\sc CASA} clean algorithm in the multi-frequency synthesis mode. We produced images over a wide range of resolutions and with different uv-tapers and weighting schemes to emphasize the radio emission on various spatial scales. A second order Taylor polynomial \citep[{\tt nterms = 2},][]{mcmullin2007} was used, to take into account the spectral behaviour of bright sources. To refine the calibration for the target, we performed two cycles of phase and one cycle of amplitude and phase self-calibration. 

We also constructed a spectral index map using the VLA D configuration image at 1.5\,GHz and a GLEAM radio image at 158\,MHz.
To create a reliable spectral index map it is essential to compare the images using the same resolution and the same pixel size. First, we deconvolved the two-dimensional VLA calibrated data using the {\tt uvtaper} option within the {\sc CASA} {\tt clean} algorithm to create a total flux intensity image close to the resolution of the GLEAM image. Then, we smoothed the image with the exact GLEAM restoring beam size\footnote{The GLEAM restoring beam is $155.57\,\arcsec \times 143.65\,\arcsec.$} using the {\sc CASA} task {\tt imsmooth}. Secondly, the GLEAM image was regridded to get the same pixel size as the VLA D configuration image. We used the {\sc CASA} {\tt immath} task, to calculate flux densities, spectral indices, and their errors following \cite{rajpurohit2018}. We assumed flux density uncertainties ($f_{\rm err}$) of 4\,$\%$ for the VLA data \citep{perley2013} and 8\,$\%$ for the GLEAM data \citep{hurley-walker2017}.

\begin{figure*}
\centering
\includegraphics[width=249pt]{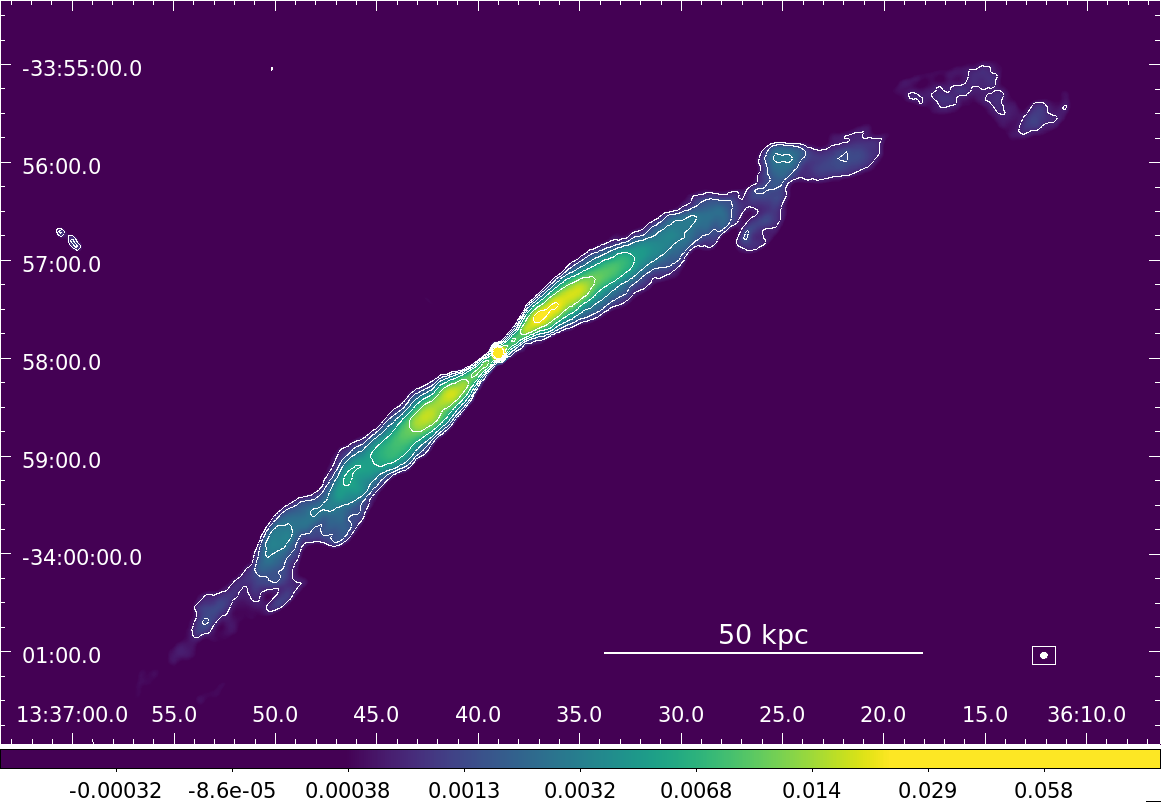}
\includegraphics[width=222pt]{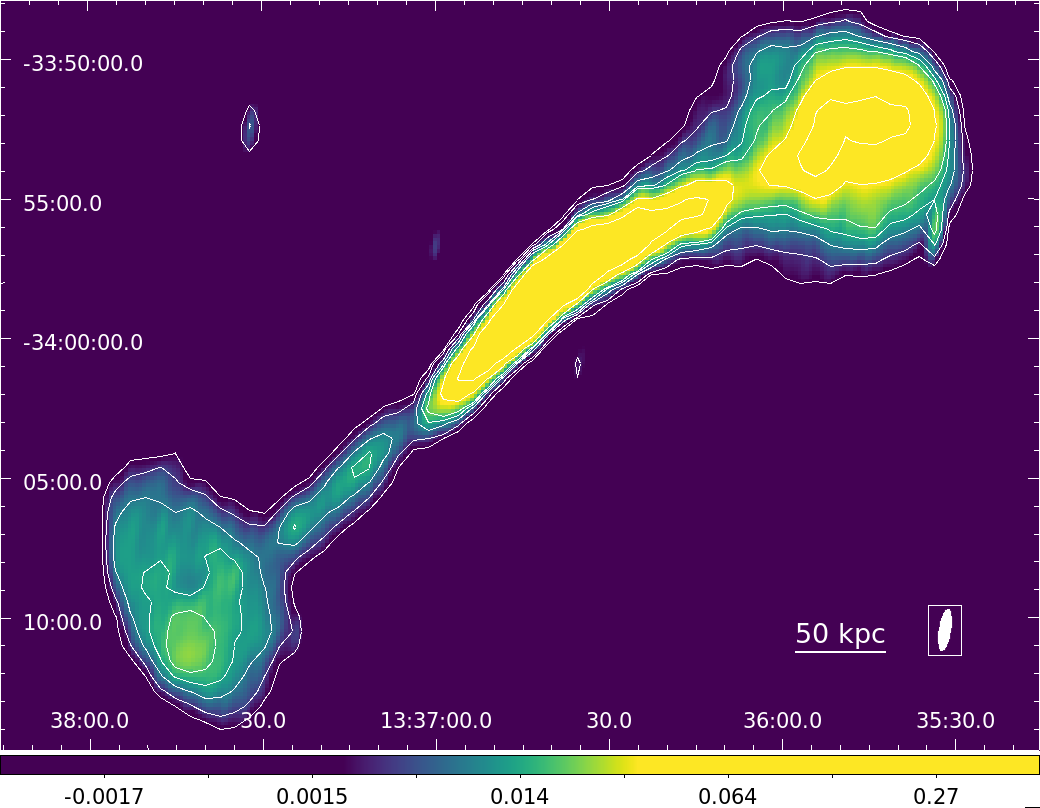}
\caption{The VLA total intensity images of IC\,4296 at 1.5\,GHz. {\it Left panel}: the A configuration image shows that the surface brightness of jets decreases as a function of radius, although we also see re-brightening in the radio plumes at $\sim 10\,\rm{kpc}$ from the nucleus. 
The restoring beam size is $3.94\,\arcsec \times 3.26\,\arcsec$
and is shown in the bottom right corner. 
{\it Right panel}: the D configuration image revealed well-confined radio lobes extending up to almost 300\,kpc from the central regions of IC\,4296, previously reported by \citet{killeen1986I}. 
 The restoring beam is $90.15\,\arcsec \times 23.02\,\arcsec$ and is shown in the bottom-right corner. The contour levels are at [-1,1,2,4,8,16] $\times$ 5\,rms noise, with corresponding rms noise of $0.06\,\rm{mJy}\,\rm{beam}^{-1}$ and of $0.2\,\rm{mJy}\,\rm{beam}^{-1}$ for VLA A and D configuration data, respectively.}
\label{fig:jvla_calibrated}
\end{figure*}

\subsection{Archival X-ray observations}
\label{sect:xrays}

\subsubsection{Archival {\it Chandra} observations and analysis}
\label{sect:chandra_observations}
IC\,4296 was observed by the {\it Chandra} Advanced CCD Imaging Spectrometer in the S-array ({\sc ACIS-S}) on 2001 September 10 (Obs. ID: 2021) and 2001 December 15 (Obs. ID: 3394) with a total clean exposure time of 48\,ks. The on-axis point-spread function (PSF) is better than 1\,arcsec for {\it Chandra}. The data reduction and analysis are described in detail in \citet{lakhchaura2018}. \\

\subsubsection{Archival {\it XMM-Newton} observations and analysis}
\label{sect:xmm_observations}
In addition to \textit{Chandra}, we also used an archival \textit{XMM-Newton} observation (Obs. ID: 0672870101) of IC\,4296, which was performed on 2011 July 11. We took advantage of both the EPIC (MOS\,1, MOS\,2, and pn) and RGS (RGS\,1 and RGS\,2) instruments to fully investigate this system. The on-axis PSF of the X-ray telescopes on \textit{XMM-Newton} is of the order of 10\,arcsec.

The entire EPIC data reduction procedure and EPIC (European Photon Imaging Camera) imaging extraction were done using the XMM Science Analysis System ({\sc SAS}) software v17.0.0 following \citet{mernier2015} to which we refer for further details.
After filtering the EPIC data from flaring events, the cleaned exposures are 46, 47, and 44 ks for MOS\,1, MOS\,2, and pn, respectively. 

The EPIC spectral analysis reported in Sect.\,\ref{sect:comp_Xray} follows the general prescription detailed in \citet{mernier2015}. We used the spectral fitting package {\sc SPEX} v3.04 \citep{kaastra1996,kaastra2017} to analyse the obtained EPIC spectra. Since the signal-to-noise ratio in the extracted EPIC regions is much weaker than in the central regions analysed with \textit{Chandra} {\sc ACIS-S} (Sect.~\ref{sect:comp_Xray}), a careful modelling of the background is necessary here. This modelling, taking into account the galactic thermal emission, the local hot bubble, the unresolved point sources (or cosmic X-ray background), the quiescent particle background, and the residual soft-proton background is also fully described in \citet[][]{mernier2015}, and references therein.
Finally, we modelled the X-ray emission of IC\,4296 with a redshifted and absorbed collisional ionisation equilibrium plasma model ({\tt cie}), in which the abundances are fixed to 0.3 proto-solar \citep{urban2017}. The MOS\,1, MOS\,2, and pn spectra were fitted simultaneously and the spectra were deprojected using the {\tt dsdeproj} tool \citep{russell2008}. The free parameters are thus the temperature ($kT$) and the SPEX normalisation.

The RGS data reduction was performed using the {\sc SAS} task \texttt{rgsproc}, which also takes care of filtering the data from flared events and of estimating the background from the RGS CCD\,9 (where no source counts are expected). The first order RGS\,1 and RGS\,2 raw spectra, extracted within 0.8\,arcmin (i.e. 90\,$\%$ of the PSF) in the cross-dispersion direction, are then background-subtracted and combined using the {\sc SAS} task \texttt{rgscombine}. 
Compared to EPIC, the RGS spectra are in principle much better suited for investigating the heating-cooling balance in the central X-ray atmosphere of IC\,4296 (Sect.~\ref{sect:comp_Xray}). Therefore, in addition to the {\tt cie} emission, we model our combined RGS spectrum with a classical `cooling-flow' ({\tt cf}) component. The multiplicative component {\tt lpro} (calculated from the MOS\,1 detector image in the 0.3--2\,keV band) is also applied to account for the instrumental broadening of the lines due to the spatial extent of the source. The lower and higher temperature limits of the {\tt cf} model are respectively fixed to 0.1\,keV and tied to the temperature of the {\tt cie} model. The free parameters are the normalization of the {\tt cie} component, the cooling rate of the {\tt cf} component, and the O, Ne, and Fe abundances. The other abundances are tied to that of Fe, with the exception of Mg which is tied to the O abundance.

\subsection{Archival optical observations and analysis}
\label{sect:opt_observations}
Narrow-band H$\alpha +$[\ion{N}{II}] imaging of IC\,4296 was obtained by Sun et al. (in prep.) with the 4.1\,m SOAR telescope using the SOAR Optical Imager (SOI). The full description of these observations is given in the survey paper (Sun et al. in prep.), but we briefly summarize them here. IC\,4296 was observed on 2015 July 21 for 1800\,s in both on (6649/76) and off (6737/76) band filters\footnote{CTIO (Cerro Tololo Inter-American Observatory) H$_{\alpha}$ filter 6649/76 ($\lambda_{\rm{cen}}$ = 6650{\AA}, FWHM = 77{\AA}) and 6737/76 ($\lambda_{\rm{cen}}$ = 6746{\AA}, FWHM = 86{\AA}).}. Basic reduction was performed using SOI-specific {\sc IRAF} scripts based on the {\tt mscred} package.

In addition to the SOAR image, we used archival {\it HST} Imaging with the Advanced Camera for Surveys (ACS) High Resolution Channel (HRC) obtained as part of program GO-9838 (PI: L. Ferrarese) observed on 2004 August 7. ACS/HRC images are in the F435W filter and the FR656N filter, which was tuned to a central wavelength of $6644.3\,\rm{\AA}$; this tuning samples H$\alpha$ at the redshift of IC\,4296. These observations were previously discussed by \citet{dallabonta2009} and presented also in \cite{lauer2005} and \cite{balmaverde2006}. We did not continuum-subtract the ACS/HRC images, but a strong nuclear source is seen in the FR656N image that is not seen in the corresponding F435W image. As was discussed by \citet{dallabonta2009}, there is strong H$\alpha +$[\ion{N}{II}] emission that extends beyond the central dust disk visible in the right panel of Fig.\,\ref{fig:opt} (for more details see also Sect.~\ref{sect:results_opt}).

\section{Results}
\label{sect:results}

\subsection{Radio morphology}
\label{sect:radio_morphology}
The 1.5\,GHz VLA A configuration radio image is shown in Fig.\,\ref{fig:jvla_calibrated} ({\it left panel}) with a resolution of $3.94\,\arcsec \times 3.26\,\arcsec$ and a root-mean-square (rms) noise of 0.06\,mJy\,beam$^{-1}$. The total intensity image clearly reveals a bright nucleus and extended emission in two jets, with several knots on both sides, terminating in diffuse radio plumes. 
The two well-collimated and almost symmetric jets emit via synchrotron radiation in the central regions. However, interestingly, the jets show noticeable re-brightening further away from the nucleus at $r\sim 10\,\rm{kpc}$, where the radio plumes broaden, as mentioned previously by \cite{killeen1988}. Based on our VLA data obtained in A configuration (which is less sensitive to extended structures), the total extent of the radio structure is $\sim 120\,\rm{kpc}$ with the northwestern jet extending out to $r\sim 70\,\rm{kpc}$ and the southeastern to $r\sim 50\,\rm{kpc}$. The total integrated flux density of the source in A configuration is $3.98\pm0.40\,\rm{Jy}$. IC\,4296 shows a gradual decrease in surface brightness from the centre, which is typical of Fanaroff-Riley (FR~I) radio galaxies. \\

The low-resolution D configuration image is shown in Fig.\,\ref{fig:jvla_calibrated} ({\it right panel}). The image has a resolution of $90.15\,\arcsec \times 23.02\,\arcsec$ and the rms noise level is $\sim 0.2\,\rm{mJy\,beam^{-1}}$. The total integrated flux density of IC 4296 in D configuration is $8.30\pm0.83\,\rm{Jy}$. The difference between the extracted integrated flux density values in A and D configurations lies in the fact that the A configuration data with the missing short baselines are not able to recover the highly peaked signal from the more extended diffuse emission in the jets and lobes.\
Two well-confined lobes are visible in the D configuration image at a projected distance of $\sim $230\,kpc and 290\,kpc from the bright nuclear region, in the northwestern and southeastern directions, respectively.

 \begin{figure}
    \centering
    \includegraphics[scale=5,width=250pt]{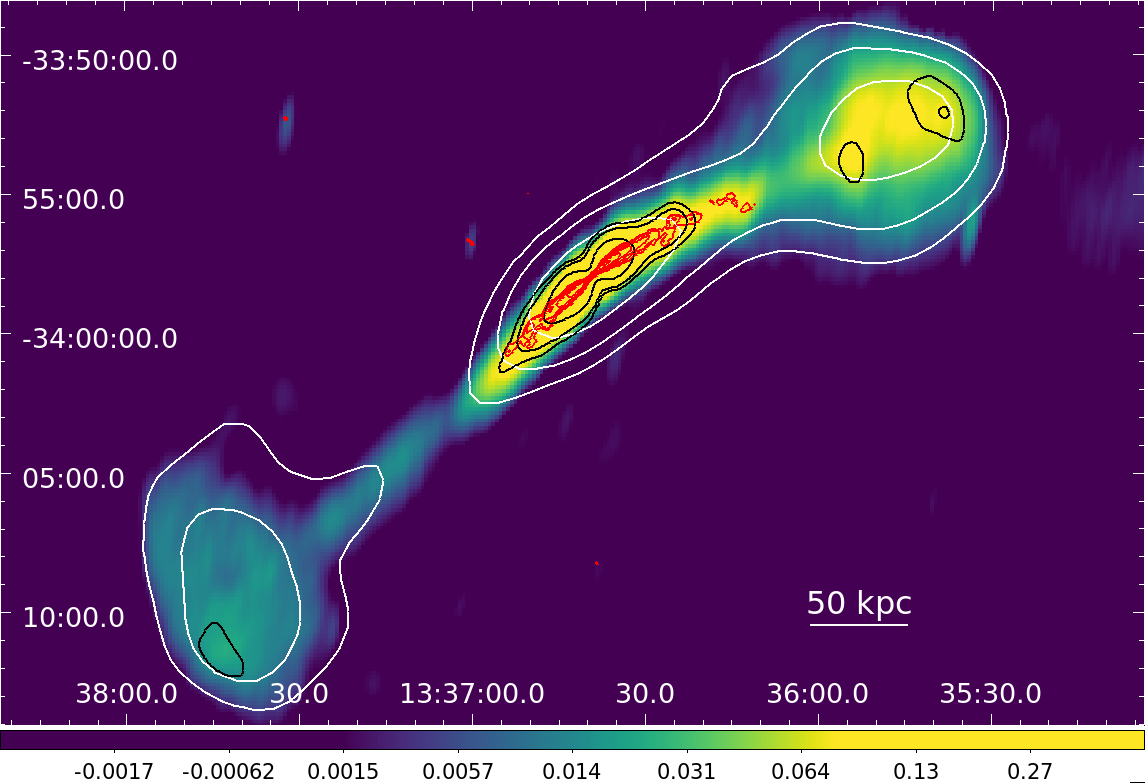}
     \caption{The VLA D configuration radio map at 1.5\,GHz overplotted with the TGSS radio contours at 153\,MHz (in black), VLA A configuration contours at 1.5\,GHz (in red), and the GLEAM radio maps at 154--162\,GHz (in white). Contours levels were created at $[-1, 1, 2, 4, 8, 16]\times5\,$ rms noise for the TGSS, GLEAM, VLA A configuration data with corresponding rms noise of 3, 0.12, and $0.06\,\rm{mJy}\,beam^{-1}$.}
    \label{fig:tgss}
   \centering
\end{figure}
Since the plumes and/or lobes in FR\,I and FR\,II radio galaxies tend to have steep radio spectra, we also inspected the archival TGSS radio maps at 153\,MHz. These revealed a radio structure close to that observed at 1.5\,GHz in our VLA A configuration image, but also a weak diffuse lobe-like emission extending far beyond the A configuration image (black contours in Fig.\,\ref{fig:tgss}). The GLEAM data at 154--162\,MHz also confirm the two extended lobe-like structures (white contours in Fig.\,\ref{fig:tgss}). 

\begin{figure}
    \centering    
    \includegraphics[width=200pt]{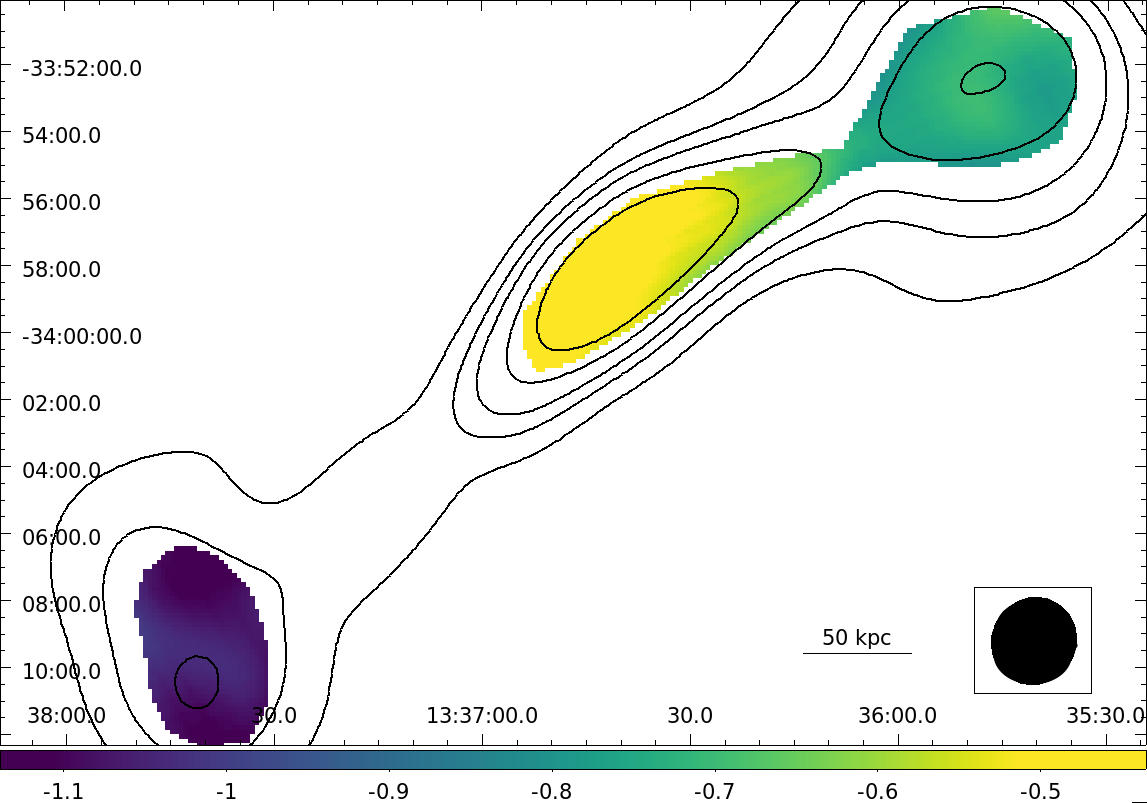}
     \caption{The spectral index map for IC\,4296 reveals a flat spectral index in the central region of the jet, whereas in the two lobes we observe a steep spectral index of $-0.7$ and $-1.1$. The map was created using the~158\,GHz GLEAM and the 1.5\,GHz VLA D configuration image at $155\,\arcsec \times 143\,\arcsec$ resolution (the beam is shown in the right corner of the map) and it is overlaid by VLA contours corresponding to the VLA smoothed image, in respect to GLEAM image to reach the same beam size. The rms noise reached value of $\sim 1.5\,\rm{mJy}\,beam^{-1}$. Contour levels are $[-1, 1, 2, 4, 8, 16]\times 5\,{\mathrm{rms\, noise}}$.}
    \label{fig:spim}
   \centering
\end{figure}

The spectral index map created using the GLEAM data at 158\,MHz and the VLA D configuration data at 1.5\,GHz is shown in Fig.\,\ref{fig:spim}.
As expected, the central regions of the galaxy show a flat spectral index of $-0.5$ and a steepening is present in the lobes. They both have steep spectra with indices of $-0.7$ and $-1.1$, respectively (Fig.\,\ref{fig:spim}). The spectral index map appears uniform over the two lobes in agreement with the lack of clear hot spots that would have flatter radio spectra. The uncertainties on the spectral index are of the order of a few percent.\

The radio power of $\sim 10^{24}\rm{\,W\,Hz}^{-1}$ at 1.4\,GHz\footnote{With corresponding flux density $\sim 8\,Jy$.} obtained from {\sc CASA} total intensity D configuration image is close to the boundary radio power of the FR dichotomy of $\sim10^{24.5}\rm{\,W\,Hz}^{-1}$ at 1.4\,GHz \citep{owen1994}. With its absolute R-band magnitude of $M_{\mathrm{R}}=-23.33\pm0.24\,\rm{mag}$, IC\,4296 is placed below the boundary division line of the Ledlow-Owen diagram, where most of the FR~I sources are located. 

\subsection{Comparison with the X-ray data}
\label{sect:comp_Xray}
 
\begin{figure*}
    \centering
    \includegraphics[width=250pt]{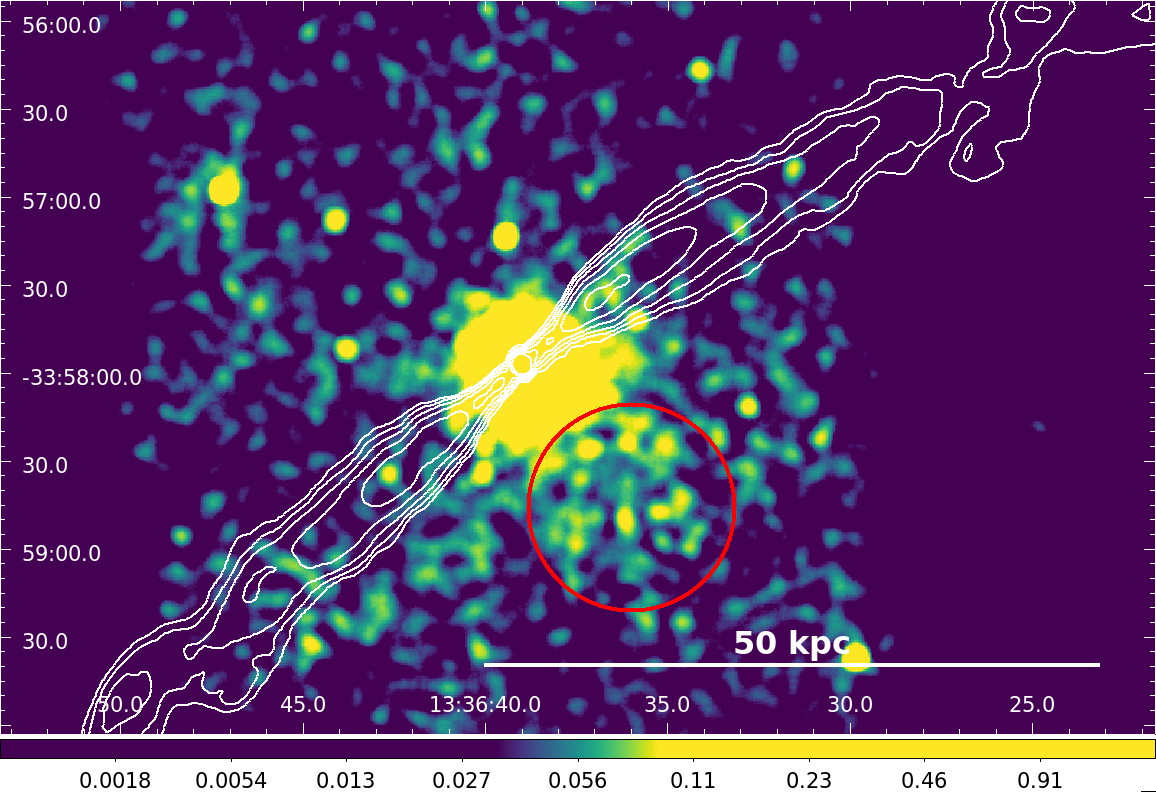}
    \includegraphics[width=219pt]{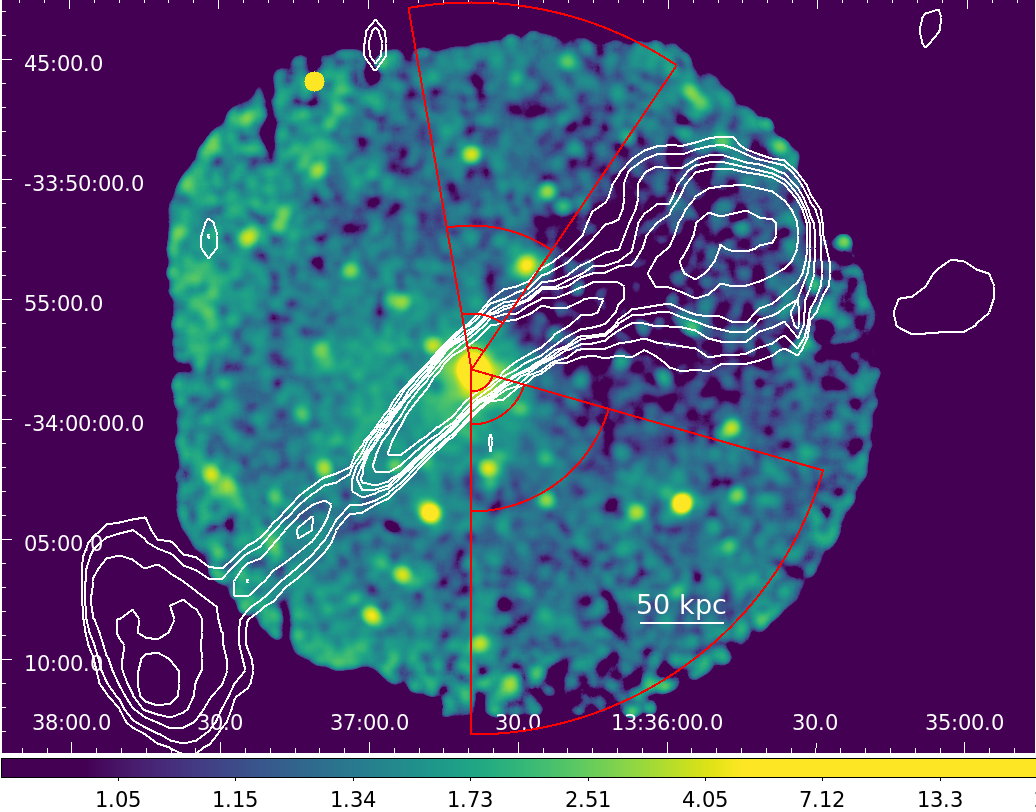}
     \caption{{\it Left panel: }The {\it Chandra} X-ray image at 0.5--5\,keV overlaid with white contours of our VLA image at 1.5\,GHz obtained in A configuration with same contour levels as defined in Fig.\,\ref{fig:jvla_calibrated}. 
     The image reveals an excess of the X-ray emission, marked by a red circle, to the southwest of IC\,4296. {\it Right panel: }The background-subtracted, exposure-corrected {\it XMM-Newton} EPIC image (combining MOS\,1, MOS\,2, and pn) in the 0.3--2\,keV band overlaid with contours of the VLA D configuration image same as in Fig.\,\ref{fig:jvla_calibrated} ({\it right panel}). 
     The X-ray image reveals a decrement, a likely cavity, corresponding to the northwestern radio lobe. The overplotted red partial annuli indicate the spectral extraction regions used to determine the temperature and the density of the ICM outside of the northern radio lobe.}
    \label{fig:seemission}
   \centering
\end{figure*}

The characteristic radius of the X-ray atmosphere in Fig.\,\ref{fig:seemission} ({\it left panel}) observed by {\it Chandra} {\sc ACIS-S} \citep[see also][]{pellegrini2003} is much smaller than the total extent of the radio jets. The jets emanating from the nuclear region are at first well-collimated and, after piercing through the innermost parts of the hot X-ray emitting atmosphere, at $r\sim\,10\,\rm{kpc}$, they widen significantly, as previously noticed by \cite{killeen1988}.\

\begin{figure}
    \centering
   \includegraphics[scale=5,width=220pt]{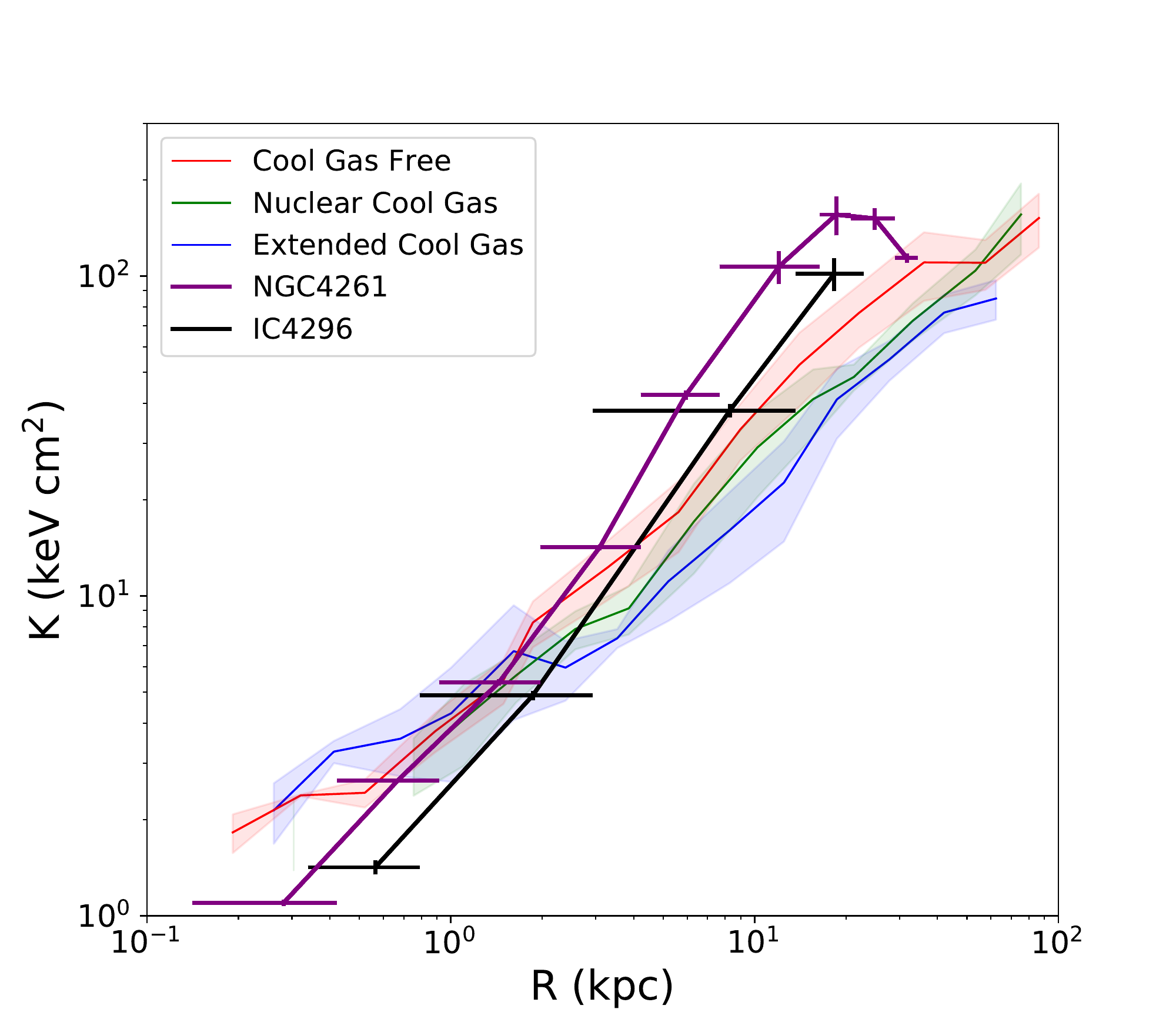}
   \caption{The entropy profiles of IC\,4296 and NGC\,4261 in comparison with the sample of \citet{lakhchaura2018}. The red, green and blue solid lines show median profiles for cool gas free, nuclear cool gas and extended cool gas systems, respectively and the shaded regions show the median absolute deviation (MAD) spreads about the medians. The entropy profiles of IC\,4296 and NGC\,4261 show much steeper trends than the other galaxies in the sample.}
     \label{fig:compIC4296_restSample}
\end{figure}

\cite{lakhchaura2018} investigated the radial profiles of the entropy and the cooling time ($t_\mathrm{cool}$) over free-fall time ($t_\mathrm{ff}$) ratio. The cooling time and entropy decrease with decreasing radius down to $<1\,\rm{kpc}$. The power-law fit of the radial entropy profile reveals a steep slope with an index of $\sim 1.2$. Even below 2\,kpc, the slope of the entropy profile remains steep with an index of $\sim 1.0$, and in the innermost radial bin the entropy reaches a value of $K_{0}\sim1.5\,\rm{keV\,cm^2}$ \citep[see the appendix in][]{lakhchaura2018}. In fact, IC\,4296 has the steepest entropy profile of all galaxies in the sample of \citet{lakhchaura2018}, followed by that of NGC\,4261 (see Fig.\,\ref{fig:compIC4296_restSample}). This will be discussed further in Sect.~\ref{sect:discussion_feedback_cycle}.\\

The cooling time in the centre of IC\,4296 is $t_{\rm cool}<11.2\,\rm{Myr}$. The ratio of the cooling time to free-fall, time $t_{\mathrm{cool}}/t_{\mathrm{ff}}$, reaches a value close to 10 in the centre of the galaxy, where the atmosphere most likely continues to cool, feeding the central supermassive black hole and powering the jets.

The archival {\it Chandra} X-ray image reveals an excess of X-ray emission in the southwestern part of the structure, in between the northwestern and southeastern side of the jet, as marked by a red circle in the {\it Chandra} X-ray image in Fig.\,\ref{fig:seemission} ({\it left panel}). The detection significance of such X-ray diffuse emission is above $5\,\sigma$.
Surface brightness profiles derived along this extended X-ray emission do not reveal any sharp edges.

Our {\it XMM-Newton} EPIC image extracted in the 0.3--2.0\,keV band reveals a cavity with a diameter of $\sim 160\,\rm{kpc}$ associated with the northwestern radio lobe (see Fig.\,\ref{fig:seemission}; {\it right panel}). We find a small misalignment between the radio lobe and the X-ray cavity. It could be the result of projection effects, where the part of the cavity closer to the mid-plane of the source is clearly visible in the X-ray image, while the part associated with the visible radio emission might be offset from the mid-plane and thus not seen clearly in the images. \

\begin{figure}
    \centering
   \includegraphics[scale=5,width=220pt]{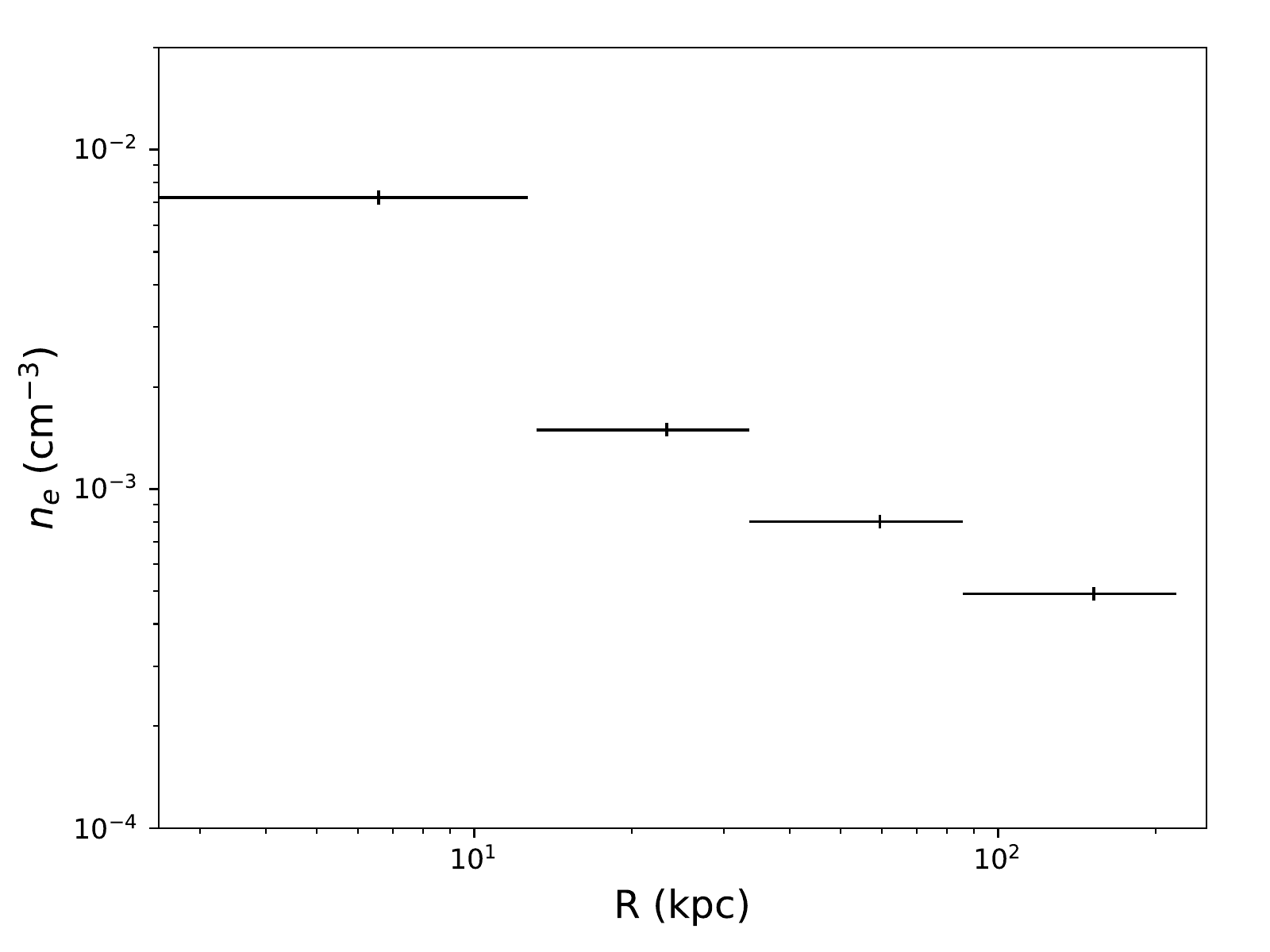}
   \caption{Deprojected electron density profile for the gas surrounding the northwestern cavity of IC\,4296. The two spectral extraction regions, whose values have been combined and averaged in this figure, are shown by the red sectors in Fig.~\ref{fig:seemission} ({\it right panel}).}
     \label{fig:density_profile}
\end{figure}

We extracted and fitted the deprojected spectra from four partial spherical shells in the direction of the northwestern jet (traced by the red sectors in Fig.~\ref{fig:seemission} {\it right panel}). The corresponding radial profile of the electron density ($n_e$) is shown in Fig.~\ref{fig:density_profile}. We have subtracted the emission contribution of the gas at larger radii (i.e. beyond our field of view), which we estimated by extrapolating a beta model fit to our initial density profile. The temperature and total density (electrons plus ions) of the gas surrounding the cavity (i.e. the last radial bin in Fig.\,\ref{fig:density_profile}) are respectively $kT=0.95\pm0.01\,\rm{keV}$ and $n=9.6\pm0.05\times10^{-4}\,\rm{cm}^{-3}$, which corresponds to a pressure of $nkT=9.12\times10^{-4}\,\rm{keV}\,\rm{cm}^{-3}$. Assuming the northwestern cavity is spherical with a radius of 80\,kpc, the $4pV$ work performed by the jet to displace the ICM is $\sim3.7\times10^{59}\,\rm{erg}$. Assuming a similar enthalpy for the southeastern cavity, which is outside of the {\it XMM-Newton} field-of-view, we obtain a total enthalpy of $\sim7\times10^{59}\,\rm{erg}$.\

\begin{figure}
    \centering
   \includegraphics[scale=5,width=250pt]{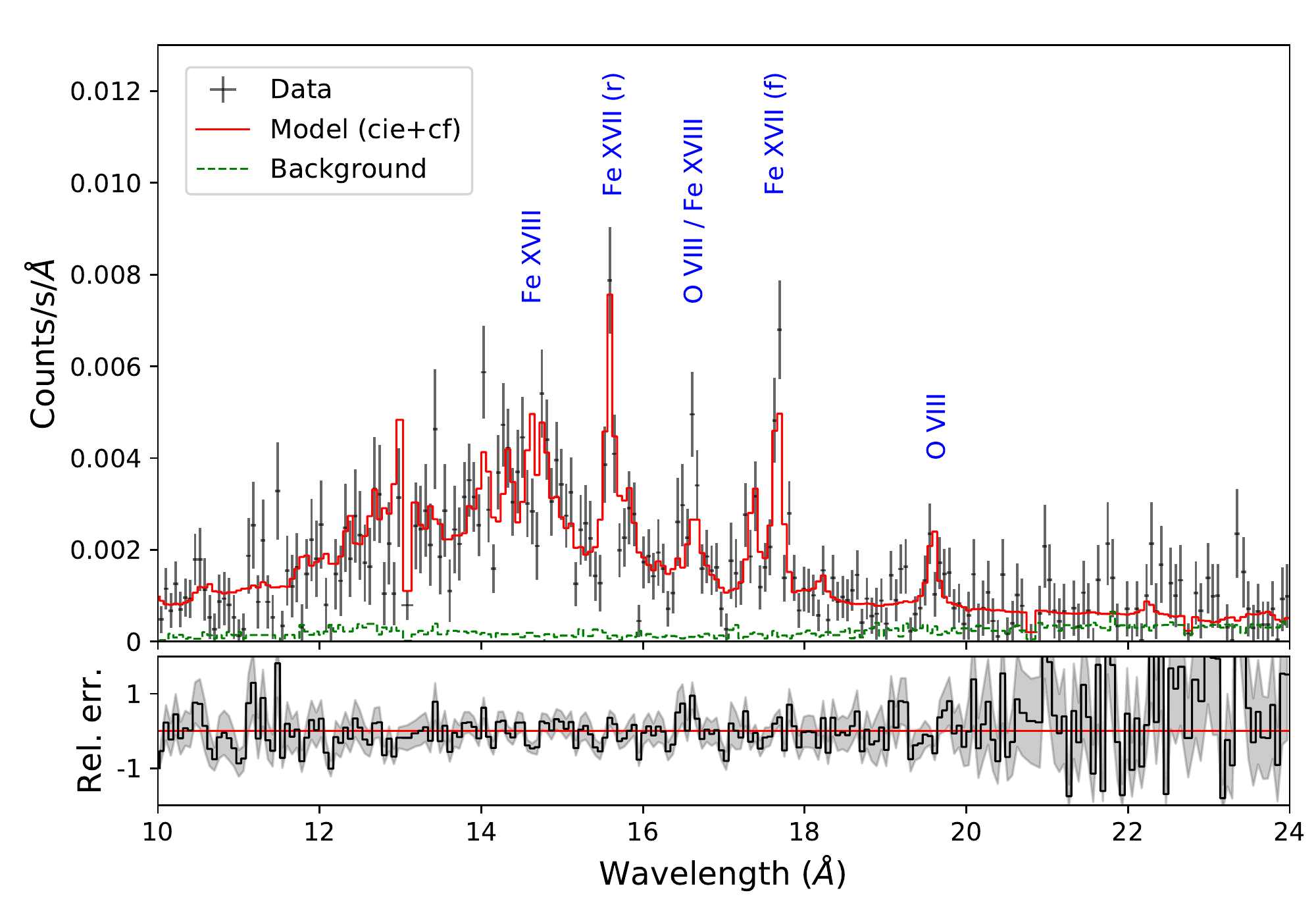}
   \caption{Combined first order RGS\,1+RGS\,2 spectrum of IC\,4296, extracted within 0.8\,arcmin in the cross-dispersion direction. Our best-fit model ({\tt cie+cf}, see text) is overplotted in red.}
     \label{fig:RGS}
\end{figure}

The combined {\it XMM-Newton} RGS\,1+RGS\,2 spectrum is shown in Fig.~\ref{fig:RGS}. Some lines (e.g. Fe\,XVII at 15\AA~and 17\AA, O\,VIII at 19\AA, rest frame) are clearly identified as they are typical of a cool ($\lesssim$1 keV) gas. Our best-fit model provides a temperature of $0.79 \pm 0.05\,\rm{keV}$ and an Fe abundance of $0.13 \pm 0.02$ for the {\tt cie} component, as well as a cooling rate of $4.5 \pm 1.0$ $M_\odot$ yr$^{-1}$ for the {\tt cf} component. In this spectrum of limited quality, however, the cooling rate and Fe abundance parameters seem to share some degeneracy, with our inferred cooling rate dropping to $2.1 \pm 0.6$ $M_\odot$ yr$^{-1}$ when fixing the abundances to 0.3 proto-solar. This is further discussed in Sect.~\ref{sect:discussion_feedback_cycle}. \

\subsection{Comparison with the optical data}
\label{sect:results_opt}
\begin{figure*}
\begin{minipage}[h]{230pt}
\centering
\includegraphics[width=230pt]{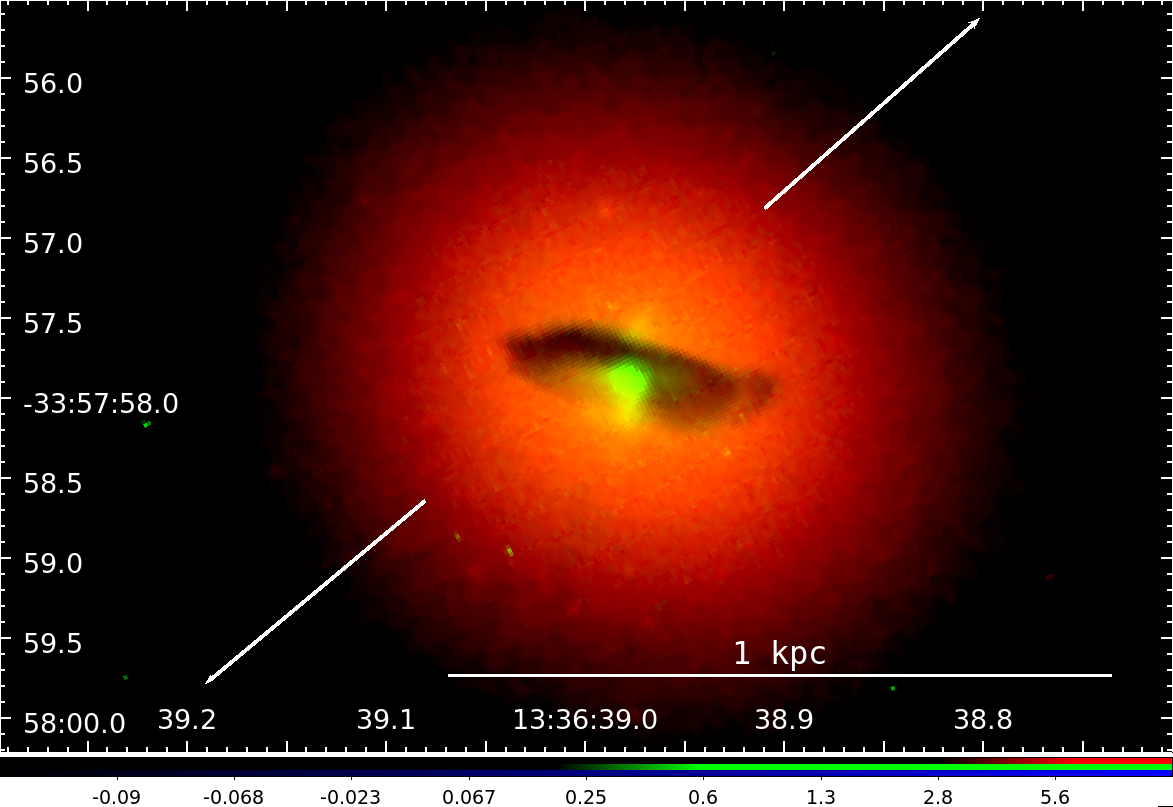}
\end{minipage}
\centering
\hspace{0.3pt}
\begin{minipage}[h]{230pt}
\centering
\includegraphics[width=230pt]{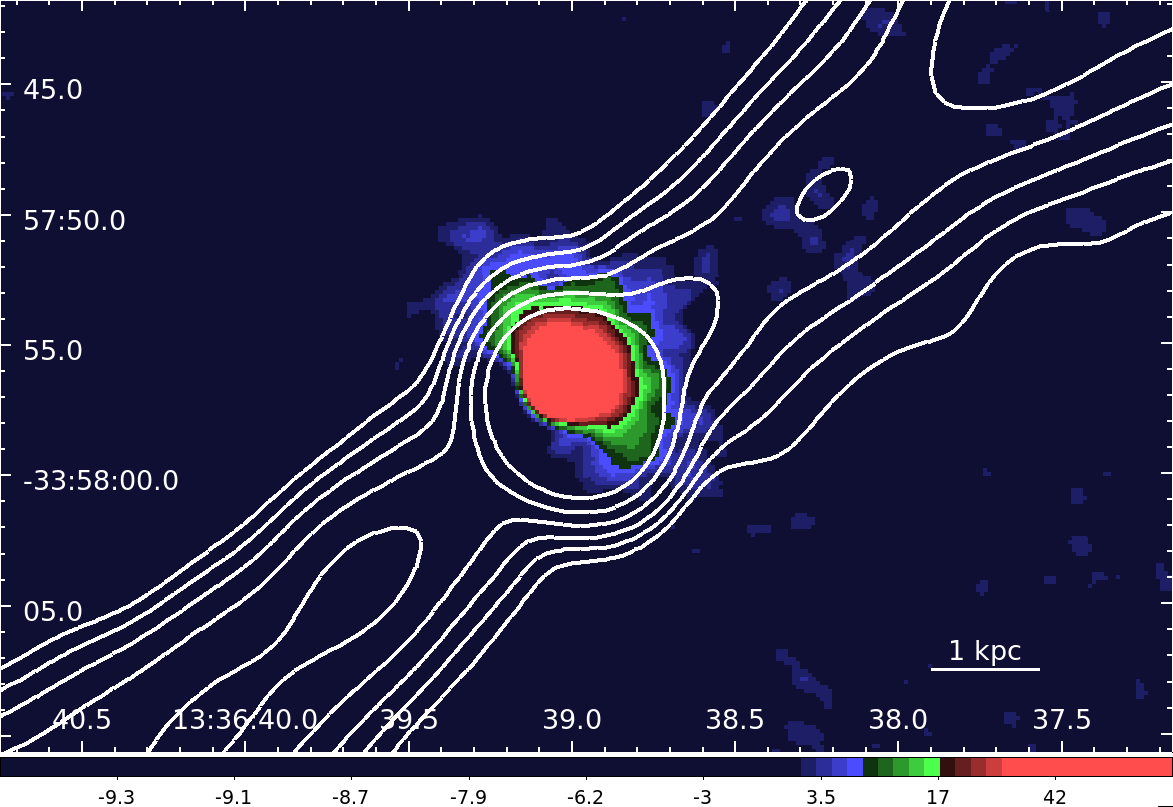}
\end{minipage}
\caption{{\it Left panel:} The {\it HST} image of a dust disk in IC\,4296. The narrow band H$\alpha$ image (in green) was obtained using ramp filter adjusted to the redshift of the galaxy. Two white arrows represent jets emitted from the central nuclear regions.
{\it Right panel:} The narrow band H$\alpha +$[\ion{N}{II}] image from the SOAR telescope overlaid with the VLA A configuration radio contours.}
\label{fig:opt}
\end{figure*}

The narrow band SOAR H$\alpha +$[\ion{N}{II}] image shown in Fig.\,\ref{fig:opt} ({\it right panel}) indicates that the ionized gas is circumnuclear and the ionized disk is perpendicular to the jets. The SOAR data also provide a hint for entrainment along the northwestern radio plume. The {\it HST} data show a warped, dust disk with a $0.9\,\rm{\arcsec}$ radius (see Fig.\,\ref{fig:opt}; {\it left panel}), which does not appear to be perpendicular to the jets \citep[see][for more details]{schmitt2002} and is visible on smaller scales than the emission in the SOAR image (Fig.\,\ref{fig:opt}; {\it right panel}). Interestingly, recent studies of CO emission for a sample of giant elliptical galaxies by \citet{boizelle2017} and \citet{ruffa2019} revealed the presence of a~circumnuclear molecular gas disk, which is co-spatial with the dust \citep{ruffa2019}.
 
\section{Discussion}
\label{sect:discussion}

\subsection{The nature of the radio source}\label{sect:discussion_nature_source}
Our new VLA data in D configuration at 1.5\,GHz confirm the presence of the two lobes previously detected by \citet{killeen1986I}. While the extended emission resembles the radio morphology of FR II radio galaxies, the lobes lack compact hot spots. The total extent of the radio structure is $\sim500\,\rm{kpc}$, which is in agreement with \citet{killeen1986I}. The compactness of the radio lobes relative to their distance from the galaxy core, make IC4296 very similar to other powerful radio galaxies.

From the X-ray analysis, we conclude that it would take $1.5\times10^{8}\,\rm{yr}$ to inflate the northwestern X-ray cavity with a radius of $\sim80\,\rm{kpc}$ at the sound speed of $c_{\rm s}=516\rm{\,km\,s}^{-1}$ for $kT=1\,\rm{keV}$ gas. Assuming this expansion time, we obtain a jet power of $1.6\times10^{44}$\,erg\,s$^{-1}$. Another estimate considers the sound-crossing time of the projected distance of the cavity from the centre of the galaxy of $2.2\times10^{8}\,\rm{yr}$ resulting in a jet power of $10^{44}$\,erg\,s$^{-1}$.

The radio morphology of the source indicates possible supersonic expansion. In the boundary layer, where the well-defined northwestern radio lobe meets the ambient ICM (Fig.\,\ref{fig:jvla_calibrated}; {\it right panel}), the surface brightness is increased and the end of the jet forms a lobe with a sharp-edged morphology. According to 3-dimensional hydrodynamic simulations by \cite{massaglia2016} and \cite{perucho2007}, this sharp head of the jet is consistent with bow shocks, which are typical indicators of the supersonic expansion relative to the surrounding medium. The inconsistency with the classical FR\,I class, low power radio sources with subsonic expansions, lies in the re-brightenings and sharp-edged structure of the northwestern radio lobe suggesting its supersonic expansion.

The radio morphology of IC\,4296 shows similarities with the morphology of 3C\,348 \citep[Hercules A, often classified as a~FR\,I/II radio source;][]{siebert1996,gizani2003}, which is one of the most luminous radio sources with a~total power comparable to the Cygnus A~galaxy \citep{gizani2005}. Hercules\,A also has missing compact hotspots and possesses an unusual morphology dominated by jets and inflated steep spectrum lobes.\

The jets, piercing through the galactic atmosphere and depositing their energy into the surrounding intra-cluster medium, are also present in the FR~I~radio source NGC\,4261 \citep[a.k.a. 3C\,270;][]{o'sullivan2005}. Interestingly, we find that the derived enthalpy for the cavities/lobes of IC\,4296 is $\sim30$ times larger than for NGC\,4261 \citep{osullivan2011}, mainly due to the size of the radio lobes ($\sim6$ times larger in the case of IC\,4296).\
As further discussed in Sect.~\ref{sect:discussion_feedback_cycle}, in these two cases the central regions also show steep entropy (see Fig.\,\ref{fig:compIC4296_restSample}) and cooling time profiles \citep{werner2012,voit2015} and a~prominent nuclear dust disk fueling the AGN closely related to powerful radio jets \citep{jaffe1993,jaffe1994}, which might have cooled from the hot X-ray emitting atmosphere of the
galaxy. Such a dust and molecular disk are also present in the brightest cluster galaxy of Hydra A, which also displays powerful jet activity \citep{mcnamara1995,taylor1996}.

\subsection{The unbalanced cooling in IC\,4296}\label{sect:discussion_feedback_cycle}
The X-ray image overlaid with the radio contours, in Fig.\,\ref{fig:seemission}, indicates that while in the inner part of the galaxy, at $r\lesssim10\,\rm{kpc}$, the radio jets are well-collimated by the thermal pressure of the hot atmosphere, their width increases at larger radii. Nevertheless, these powerful radio jets pierce through the inner hot X-ray emitting atmosphere of the galaxy and extend to radii over $r\gtrsim 230$\,kpc, where they deposit their energy into the outer galactic atmosphere and the ambient ICM.

Similarly to NGC\,4261, the entropy profile of IC\,4296 (Fig.\,\ref{fig:compIC4296_restSample}) shows a steeper slope than for all the other giant elliptical galaxies analysed by \citet[][]{lakhchaura2018}. In fact, it appears to be decreasing all the way to $r<1\,\rm{kpc}$ \citep[see also][]{panagoulia2014,babyk2018}. At $r<1\,\rm{kpc}$, $t_{\rm cool}/t_{\rm ff}$ drops to $\approx$10 and the gas may become thermally unstable \citep{sharma2012,gaspari2012,kunz2012,mccourt2012,gaspari2013,voit2015b}. These results suggests that, whereas the expanding radio jets and lobes may be able to heat the surrounding gas at $r\gtrsim10$\,kpc, the inner, well-collimated, part of the jets do not appear to be depositing sufficient heat to the innermost rapidly cooling X-ray atmosphere.

 At first glance, this picture is in line with our analysis of the \textit{XMM-Newton} RGS spectra, as they are consistent with an appreciable mass deposition rate of $4.5\pm1.0\,M_\odot\,\rm{yr}^{-1}$. This rate is particularly high compared to other elliptical galaxies and groups whose RGS spectra reveal mass deposition rates typically less than 1 $M_\odot$~yr$^{-1}$ \citep{liu2019}. We caution, however, that this measurement suffers from large systematic uncertainties because the limited quality of the data implies degeneracies between some parameters. For example, fixing all the abundances to 0.3 proto-solar \citep[which is more realistic than our best-fit 0.13 proto-solar value; e.g.][]{urban2017} yields a lower mass deposition rate of $2.1\pm0.6\,M_\odot\,\rm{yr}^{-1}$. Moreover, we cannot exclude even higher abundances (i.e. close to the proto-solar value) in the inner $\sim$10 kpc region, which would then further revise lower the inferred cooling rate. Future deeper observations will thus be important to confirm (or disprove) the unusually high cooling rate in this source.
 Assuming nevertheless our initial deposition rate and that the bulk of this material accretes onto the central supermassive black hole and is converted into jets with a 10\,\% efficiency, it will result in a jet power of $\sim 2.6\times\,10^{46}\,\rm{erg\,s^{-1}}$. This is more than two orders of magnitude higher than the actual jet power estimated in Sect.~\ref{sect:discussion_nature_source} ($\sim 10^{44}\,\rm{erg\,s^{-1}}$). We therefore conclude that the currently assumed rate of cooling is largely sufficient to effectively produce the observed radio jets.
 
In the centre of the galaxy, a disc of ionized and molecular gas \citep{boizelle2017, ruffa2019} is observed, which is also visible in the {\it SOAR} image in the right panel of Fig.\,\ref{fig:opt}. The circumnuclear dust disc seen by {\it HST} in the right panel of Fig.\,\ref{fig:opt}., could be covering the central engine of the galaxy, offering a possible explanation for the apparent strongly sub-Eddington nuclear luminosity of $L_\mathrm{bol}/L_\mathrm{Edd}$\footnote{The Eddington luminosity for IC\,4296 is $\sim$1.2$\times\,10^{47}\,\rm{erg\,s^{-1}}$.}$\sim$2$\times10^{-5}$ \citep{pellegrini2003}. The true Eddington ratio necessary to explain the acceleration of jets penetrating through the ambient medium could be significantly higher.
On the other hand, the low Eddington ratio could also be a~result of an advection-dominated accretion flow (ADAF), where the heat generated via viscous dissipation in the disk is ``advected'' inwards with the accreting gas \citep{narayan1994}.

\subsection{Peculiar bending jets in a hot atmosphere}

The radio and X-ray morphologies of IC\,4296 resemble those of some other systems, such as NGC\,1265, where the powerful radio jets penetrate through the atmosphere of hot gas bound to the host galaxy, with little impact on the X-ray gas \citep{sjj05}. In NGC\,1265, the two-sided jet emerges from the atmosphere into the ICM of the Perseus cluster. Since NGC\,1265 is moving at high speed relative to the ICM, we see bright radio knots in its jet where they impinge on the ICM and are deflected \citep{odea1987}. Although IC\,4296 differs in important respects from NGC\,1265, we cannot exclude that the observed re-brightening in its jet at $r\sim$10\,$\rm{kpc}$ could have a similar origin.

The presence of a possible X-ray tail seen in the {\it Chandra} image and the location of IC\,4296 in the centre of Abell 3565 is consistent with the sloshing of the ambient ICM in the potential well of the cluster \citep[see][]{markevitch2007}. Given the relative motion, the radio jets emerging from the central atmosphere run into the ICM and are deflected in the direction opposite to the galaxy's travel, as also noted by \cite{kemp1994}. Assuming that the jets are supersonic, they could be deflected in a succession of inclined weak shocks or via magnetohydrodynamical instabilities \citep{massaglia2019}, converting jets' kinetic energy to thermal energy. Alternatively, the brightening could result from the dissipation of turbulence driven by the interaction between the jets and the external medium \citep{burns1998}. In either case, the additional thermal energy will raise the pressure (and density) in the jet plasma, causing it to expand \citep[e.g.][]{gizani1999}. This would account for the increase in the width of the radio source in the regions $\sim$10\,$\rm{kpc}$ from the central AGN, also noticed by \cite{killeen1988}.

The apparent curvature of the jets is rather weak and we may be viewing the system from a direction nearly parallel to the velocity vector. The location of the tail and the radio source curvature are both consistent with a projected relative velocity towards the northeast. 

\section{Summary and conclusions}
\label{sect:summary}
We present the results of new VLA radio observations of the giant elliptical galaxy IC\,4296, obtained in the A and D configurations. Combined with complementary X-ray and optical data, we provide a comprehensive, multi-wavelength analysis of the AGN feedback and jet-ICM interaction at play in this peculiar system. Our results can be summarised as follows:

\begin{itemize}
\item The new high-resolution VLA data in A configuration reveal the bright central region of IC4296 and two almost symmetrical knotty radio jets extending far beyond the host galaxy.
With the VLA D configuration data, we confirm the presence of well-defined radio lobes first reported by \citet{killeen1986I}. The lobes, without significant hot spots, have large diameters of $\sim$160\,$\rm{kpc}$ at radii $r\gtrsim$230\,$\rm{kpc}$ and are undetected in the higher resolution A configuration VLA images. 
\item The {\it XMM-Newton} data reveal the presence of an X-ray cavity associated with one of the large radio lobes indicating a remarkable total enthalpy $\sim$7$\times10^{59}\,\rm{erg}$. For comparison, this enthalpy is 30 times larger than for NGC\,4261, due to the 6\,times larger size of the radio lobes.
\item The observed radio morphology is consistent with 3D hydrodynamical simulations \citep[e.g.][]{massaglia2016} of supersonically expanding jets/lobes, with bow shocks created at the boundary layer of the radio lobes interacting with the ambient ICM.
\item The jets are bent likely due to the relative motion of the ICM with respect to IC\,4296 and magneto-hydrodynamic instabilities, advecting the plasma in long narrow plumes. The relative motion of the galaxy with respect to the ICM is also supported by a possible X-ray tail seen in the {\it Chandra} image. 
\item The powerful jets piercing through the hot galactic atmosphere do not appear to be depositing sufficient heat in the innermost gas to prevent it from cooling, with the bulk of their energy deposited well beyond 10\,kpc from the nucleus. The atmosphere of the galaxy can therefore continue to cool rapidly, feeding the central supermassive black hole and powering the jets. This is supported by the steep entropy and cooling time profiles of the hot galactic atmosphere which continue to drop all the way to the nucleus, where a warm H$\alpha +$[\ion{N}{II}] nebula and a cold molecular CO disk are clearly seen. As for a handful of other examples, such as NGC\,4261, the cooling continues despite the powerful jet activity.
\end{itemize}

\section*{Acknowledgements}
This work was supported by the Lend\"ulet LP2016-11 grant awarded by the Hungarian Academy of Sciences.

The National Radio Astronomy Observatory is a facility of the National Science Foundation operated under cooperative agreement by Associated Universities, Inc. 
The scientific results reported in this article are based on observations made by the {\it Chandra X-ray Observatory} and published previously in cited articles. 

Based on observations obtained at the Southern Astrophysical Research (SOAR) telescope, which is a joint project of the Minist\'{e}rio da Ci\^{e}ncia, Tecnologia, Inova\c{c}\~{o}es e Comunica\c{c}\~{o}es (MCTIC) do Brasil, the U.S. National Optical Astronomy Observatory (NOAO), the University of North Carolina at Chapel Hill (UNC), and Michigan State University (MSU). K.G. was supported by the J\'anos Bolyai Research Scholarship of the Hungarian Academy of Sciences.

This research has made use of resources provided by the Compagnia di San Paolo for the grant awarded on the BLENV project (S1618\_L1\_MASF\_01) and by the Ministry of Education, Universities and Research for the grant MASF\_FFABR\_17\_01.

This investigation is supported by the National Aeronautics and Space Administration (NASA) grants GO4-15096X, AR6-17012X and GO6-17081X.

This work is supported by the ``Departments of Excellence 2018 - 2022’’ Grant awarded by the Italian Ministry of Education, University and Research (MIUR) (L. 232/2016). 

F.Massaro acknowledges financial contribution from the agreement ASI-INAF n.2017-14-H.0. A.C.F. acknowledges support from ERC Advanced Grant 340442. K.R. acknowledges financial support from the ERC Starting Grant ``MAGCOW'', no. 714196 and M. S. acknowledge the support from the NSF grant 1714764.


\bibliographystyle{mnras}
\bibliography{clusters}


\bsp
\label{lastpage}

\end{document}